# Accurate Estimation of a Coil Magnetic Dipole Moment

Antonio Moschitta, Alessio De Angelis, Francesco Santoni, Marco Dionigi, Paolo Carbone
Department of Engineering
University of Perugia
Perugia, Italy
{antonio.moschitta,alessio.deangelis,francesco.santoni, marco.dionigi, paolo.carbone }@unipg.it

Guido De Angelis
Regione Umbria
Perugia, Italy
ing.guidodeangelis.gmail.com

*Abstract*— **In this paper, a technique for accurate estimation of the moment of magnetic dipole is proposed. The achievable accuracy is investigated, as a function of measurement noise affecting estimation of magnetic field cartesian components. The proposed technique is validated both via simulations and experimentally.**

## I. INTRODUCTION

The Internet-of-Things (IoT) paradigm is based on the development of intelligent systems, capable of collecting, aggregating, and processing information originating in the real world. As such, positioning systems (PSs) are a strong enabler for IoT based applications, that include Location Based Services, Domotics, Wireless Sensor Networks, and production line traceability. Solutions proposed in the literature are based on various measurement principles and processing techniques [1-13]. Apart from solutions based on image processing, PSs are usually based on the transmission of known signals between a mobile node and a set of beacons. Then, by measuring a set of physical quantities that depend on the transmitted signal, ranging and positioning can be performed using fitting techniques on a known propagation model. Among PSs, those based on measurement of AC magnetic fields, generated by either mobile nodes or fixed beacons are often mentioned in the recent literature, because this approach is both easily implemented and robust to most environmental factors, such as the presence of obstacles and the static geomagnetic field [7]-12].

The accuracy of positioning systems is limited by the accurate knowledge of the references and of the mobile node characteristics. For AC Magnetic PSs (MPSs) the references are often a set of fixed coils acting as beacons, described by their geometrical characteristics, position, and bearing, while the mobile node is realized by an additional coil. Hence, accurate AC MPSs require accurate knowledge of coils properties, that can be obtained either by careful manufacturing, leading to strict tolerance requirements, or by accurate coils' characterization. In short-range MPS systems applications, estimating the position with sub-centimeter accuracy and the attitude with 1° accuracy requires at least an equally accurate knowledge of the beacons' ones [14].

Moreover, active coils used in MPSs are often described by an approximated model, summarizing coils' knowledge by their magnetic dipole moments. In fact, the magnetic field induced in a given position by an active coil can be expressed easily, as a function of the coil's magnetic dipole moment and the vector that describes the distance between the coil and the position of interest [9][10][14]. The voltage appearing at the output of a probe coil can be expressed in a similar way. Thus, by assuming that voltage or magnetic field measurements are collected between the mobile node and a set of beacons, the position and the attitude of a mobile coil can be estimated using numerical fitting. This leads to computationally light positioning algorithms, suitable for real time applications, especially with respect to models that estimate the magnetic field using finite element analysis [15]. Moreover, the approach based on magnetic dipole moment can be indifferently applied to MPSs featuring active beacon and mobile nodes equipped with a passive probe or to MPSs with a dual architecture, featuring an active mobile node and a set of passive beacons.

Consequently, this paper is focused on a simple characterization technique, aimed at accurately estimating the magnetic dipole moment of a coil. While magnetic dipole measurement is mentioned in the literature, most works do not target accurate characterization of active coils, being mostly focused on approximate characterization of electric appliances [16-17], on characterization of permanent magnets [18], on magnetostatic characterization of space equipment [19], or are strongly focused on modeling [20]. The proposed approach is simple and computationally light, and was validated both by simulations and experimentally, using small coils compatible with short range MPSs. It is shown that the magnitude and the direction of an active coil magnetic dipole moment can be estimated with an accuracy of less than 4%.

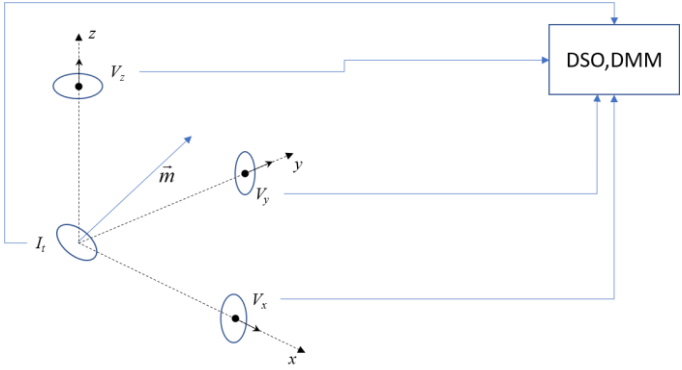

Fig. 1 – Proposed measurement setup. The magnetic dipole moment generated by a coil placed in the origin of a Cartesian coordinate system is assessed by measuring voltage at the output of a probe coil, sequentially placed on each cartesian axis and aligned to the axis itself. Phase between the phasor $I_t$ of the current feeding the active coil and each collected voltage is measured as well.

TABLE I – PARAMETERS USED IN THE NUMERICAL SIMULATIONS.

| | |
|---|---|
| Number of Monte Carlo iterations | $10^3$ |
| Number of orientations | 126 |
| Number of noise levels | 11 |
| Coil under test: radius | 5 mm |
| Coil under test: number of turns | 20 |
| Coil under test: driving current (amplitude) | 0.28 A |
| Coil under test: driving current (frequency) | 184 kHz |
| Probe coil: radius | 19 mm |
| Probe coil: number of turns | 5 |
| Sampling frequency | 3 MSa/s |
| Number of samples | $10^3$ |

## II. MEASUREMENT MODEL AND PROCEDURE

### A. Measurement model

Let us assume to operate under sinusoidal steady state, at a frequency $f_0$, and that the magnetic dipole to be measured is described by its phasor $\vec{m} \equiv (m_x, m_y, m_z)$. This magnetic dipole moment is generated by a planar and circular transmitting coil, so its magnitude $m$ is given by $m = N_t S_t I_t$, where $N_t$ is the number of coil windings, $S_t$ is the area of a coil winding, and $I_t$ is the phasor of the current stimulating the coil. Let us also assume that the transmitting coil is placed in the known position $P_m \equiv (x_m, y_m, z_m)$.

The measurement procedure can be developed by recalling that, in a given position $P \equiv (x, y, z)$, the phasor $\vec{B}$ of the AC magnetic field induced by $\vec{m}$ is given by

$$\vec{B}(x,y,z) = \frac{\mu_0}{4\pi} \frac{\left(3(\vec{m}\cdot\vec{r})\vec{r} - \vec{m}r^2\right)}{r^5}, \quad (1)$$

where $\mu_0$ is the vacuum permeability, $\vec{r}$ is the distance vector between the magnetic dipole application point $P_m$ and the position $P$, given by

$$\vec{r} = (r_x, r_y, r_z) = (x - x_m, y - y_m, z - z_m), \quad (2)$$

and $r = \|\vec{r}\|$ is the Euclidean norm of $\vec{r}$, that is the Euclidean distance between P and $P_m$.

Let us also assume that a planar coil with radius $R_p$, $N_p$ windings, and attitude described by i.e. unit vector $\vec{n}_p = (n_{px}, n_{py}, n_{pz})$, is placed in P, acting as probe. By assuming that the magnetic field is constant across the coil section, the phasor $V$ describing the probe coil output voltage is given by

$$V = 2\pi f_0 S_p N_p \left(\vec{B}(x,y,z)\cdot\vec{n}_p\right) =$$
$$= K_p \vec{B}(x,y,z)\cdot\vec{n}_p = K_p \frac{\left(3(\vec{m}\cdot\vec{r})\vec{r} - \vec{m}r^2\right)}{r^5}\cdot\vec{n}_p =$$
$$= K_p \frac{\left(3(m_x r_x + m_y r_y + m_z r_z)\vec{r} - \vec{m}r^2\right)}{r^5}\cdot\vec{n}_p, \quad (3)$$

$$K_p = \omega_0 S_p N_p \frac{\mu_0}{4\pi}, \quad S_p = \pi R_p^2, \quad \omega_0 = 2\pi f_0$$

where $S_p$ is the probe coil area.

### B. Measurement procedure

Provided that the current feeding the transmitting coil and the voltage at the output of the coil can be simultaneously measured, (3) can be used to develop a simple measurement procedure. In particular, let us assume that, without loss of generality, the transmitting coil is placed in the origin of a Cartesian coordinate system, i.e. $P_m \equiv (0,0,0)$. For instance, if the probe coil is placed on the $x$ axis, then $\vec{r} = (r_x, 0, 0) = r_x \vec{n}_x$, where $\vec{n}_x$ is the unit vector describing the direction of the $x$ axis, and (3) reduces to

$$V = \frac{K_p}{r_x^5}\left(3(m_x r_x)r_x \vec{n}_x - \vec{m}r_x^2\right)\cdot\vec{n}_p =$$
$$\frac{K_p}{r_x^5}\left(3 m_x r_x^2 \vec{n}_x - \vec{m}r_x^2\right)\cdot\vec{n}_p = \quad (4)$$
$$\frac{K_p}{r_x^3}\left(3 m_x n_{px} - (m_x n_{px} + m_y n_{py} + m_z n_{pz})\right).$$

Moreover, if the probe coil is aligned to the $x$ axis, that is $\vec{n}_p = \vec{n}_x$, (4) further reduces to

$$V = \frac{K_p}{r_x^3}(3m_x - m_x) = \frac{2K_p m_x}{r_x^3}. \quad (5)$$

Note that the probe placement and orientation leading to (5) decouples the measurement of $m_x$ from the measurement of the components $m_y$ and $m_z$. Using (5), $m_x$ is obtained as

$$m_x = \frac{r_x^3}{2K_p}V_x, \quad (6)$$

where $V_x$ indicates the voltage phasor when the probe coil is placed on the $x$ axis and aligned with it. Note that the sign of $m_x$ is given by the sign of $V_x$, that in turn depends on the relationship between the phase lag between the current feeding the transmitting coil and the probe coil output voltage. In particular, since the system operates at very low frequencies, by assuming that the $I_t$ phasor is positive and real, $V_x$ can be either real positive or real negative. By placing the probe coil on the *y* and *z* axes, each time aligned with the corresponding axis, $m_y$ and $m_z$ can be estimated as well. The measurement setup is summarized in Fig. 1. Additional phenomena affecting the measurement result, such as gain of the measurement chain and coil resonance, may be kept into account by first calibrating the system.

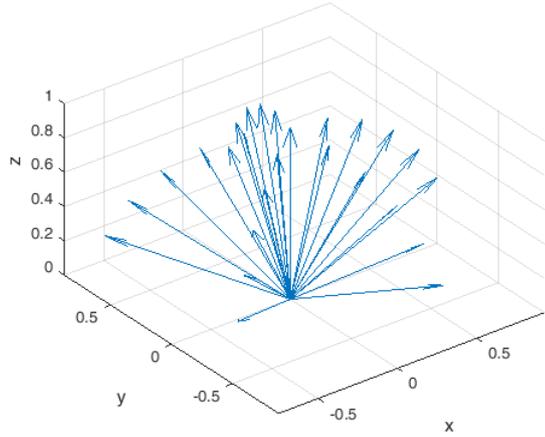

Fig. 2 – 3D representation of some of the 126 unit vectors, describing simulated orientations of the coil under test.

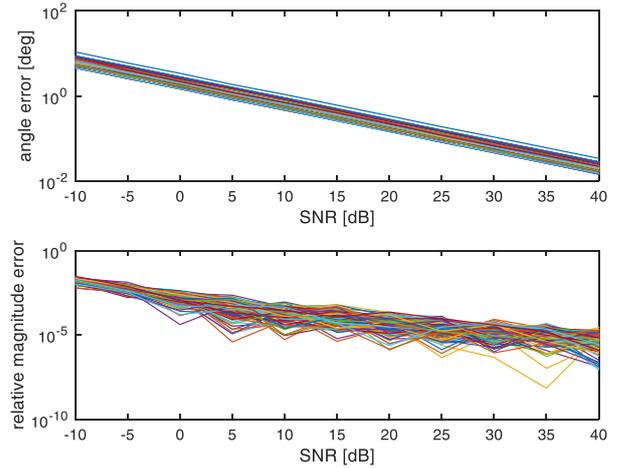

Fig. 3 – Numerical simulation results. Each curve represents one orientation of the coil under test.

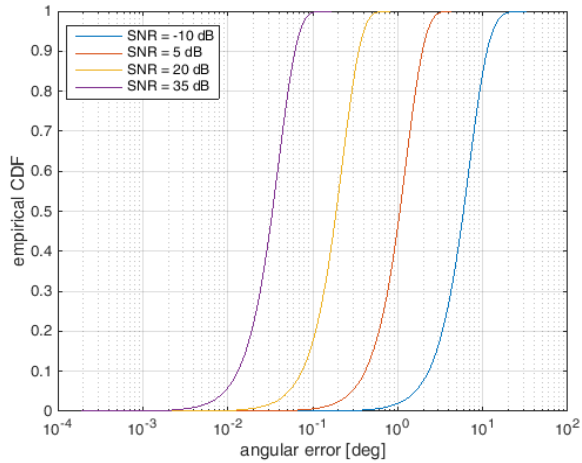

Fig. 4 – Empirical cumulative distribution function (ECDF) of the angular error.

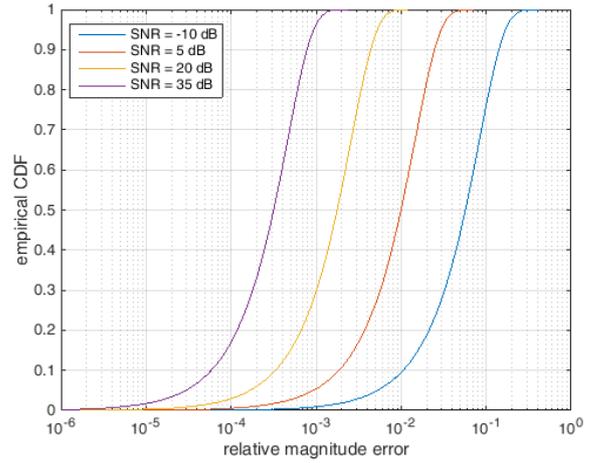

Fig. 5 – Empirical cumulative distribution function (ECDF) of the relative magnitude error

### III. SIMULATIONS RESULTS

To investigate the effect of noise and orientation on the performance of the proposed method, Monte Carlo numerical simulations were performed. The coil under test was assumed to be centered at the origin of the coordinate system with an arbitrary orientation, and to be generating a sinusoidal time-varying magnetic field. The simulations were repeated for 126 different orientations of the coil under test, with an azimuthal angle from 0° to 360° and an elevation angle from 22.5° to 90°. Some of the unit vectors describing the 126 orientations are shown in Fig. 2. Three probe coils were simulated, each centered on one of the axes, at a distance of 30 cm from the center of the coil under test, and oriented along the corresponding axis.

The voltage induced at each probe coil was simulated according to the model (1)-(5). Subsequently, additive white gaussian noise (AWGN) was added to the samples of the induced voltage and the amplitude and phase of the induced voltage were estimated based on the noisy samples using a 3-parameters sinefit algorithm [20]. Finally, the resulting $V_x$, $V_y$, and $V_z$ estimates were used to calculate the $m_x$, $m_y$, and $m_z$ components of the magnetic dipole vector, respectively, according to the proposed method as derived in Section II. The simulations were repeated for 11 different values of the noise standard deviation $\sigma_{AWGN}$, corresponding to an SNR ranging from -10 dB to 40 dB. For each value of $\sigma_{AWGN}$ and for each value of the orientation, $10^3$ Monte Carlo iterations were performed. The numerical simulation parameters are summarized in Table I.

We define the angular error as the angle between the direction of the true magnetic dipole moment and the direction of the estimated magnetic dipole moment. Moreover, the magnitude error is defined as the difference between the true magnitude of the magnetic dipole moment of the coil under test and its estimated value. Finally, we evaluate the relative magnitude error as the absolute value of the magnitude error divided by the true magnitude of the magnetic dipole moment of the coil under test.

Results for varying signal-to-noise ratio and orientations are shown in Figs. 2-5. It can be noticed that, if the SNR at the probe coil is 5 dB or greater, the proposed method allows for estimating the magnetic dipole moment vector with an average angular error of less than 1° and an average magnitude error of less than 0.1%.

## IV. MEASUREMENT RESULTS

In order to validate the proposed measurement method, experiments were performed by means of the experimental setup shown in Fig. 6. The coil under test was driven with a sinusoidal voltage signal at 184 kHz, 20 $V_{pp}$. The driving signal was provided by an Agilent 33220a function generator. A probe coil was placed at fixed and known positions. Test coil and probe coil parameters are shown in Table II. The probe coil was connected in parallel with a 330 nF capacitor (LC circuit) and the voltage gain (Q) due to the resonant circuit was measured to be 5.57. The voltage induced on the probe coil was amplified by an instrumentation amplifier (INA), AD8421 by Analog Devices, with a gain of 100. The output of the INA was connected to a Fluke 8845A voltmeter. Furthermore, the driving signal from the function generator was connected to channel 1 of an oscilloscope, while the output of the INA was connected to channel 2 of the oscilloscope. This allowed for identifying whether the driving signal and the received signal were in phase or in antiphase.

The coil under test was placed in the center of a 3D-printed holder realized as an icosahedron. The icosahedron shape was used because it allowed for placing the test coil in a set of known and controlled orientations. The center of the coil under test coincided with the origin of the local coordinate frame considered in the experiments. The probe coil was placed in the following three positions (coordinates expressed in meters), $P_x$=(0.194, 0, 0), $P_y$=(0, 0.194, 0), and $P_z$=(0, 0, 0.214). This corresponds to placing the coil center on the *x*, *y*, and *z* axes respectively. Each time, the coil axis was aligned to the corresponding cartesian axis.

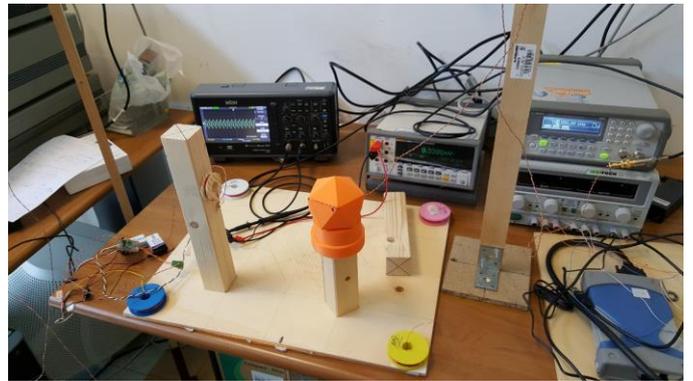

Fig. 6 – The experimental setup.

TABLE II – EXPERIMENT SETUP: TEST COIL AND PROBE COIL PARAMETERS

| Coil under test: radius | 5 mm |
|---|---|
| Coil under test: number of turns | 20 |
| Coil under test: driving current (amplitude) | 0.28 A |
| Coil under test: driving current (frequency) | 184 kHz |
| Probe coil: radius | 19 mm |
| Probe coil: number of turns | 5 |

The measurements were performed for a set of orientations of the coil under test. For each orientation and probe position, the measured voltages are shown in Table II. Subsequently, the measured voltage was used to estimate the coil dipole moment, obtaining the results shown in Table III. The measured magnetic dipole intensity is compared with the theoretical value 4.33x10-4 $Am^2$ in order to obtain the relative error reported in the rightmost column of Table III.

Note that the large azimuth error affecting the first measure (-55.95° against a true value of 0°) does not represent a measurement problem. Indeed in this case the test coil magnetic dipole is parallel to the z axis (elevation 90°), hence any azimuth defines the same direction because of the cylindrical symmetry of the problem. Thus, even in this case, the measured direction deviates from the vertical direction by an angle of only 0.81°. In all considered cases a small angular error was obtained, upper bounded by 3.4°.

TABLE III – EXPERIMENTAL RESULTS

| Test coil direction (deg) | | | | Angular error | Measured voltage (mV) | | | Magnetic dipole (10-4 $Am^2$) | |
|---|---|---|---|---|---|---|---|---|---|
| Azimuth | | Elevation | | | Probe 1 | Probe 2 | Probe 3 | Measured | Error % |
| Real | Measured | Real | Measured | | | | | | |
| 0.00 | -55.95 | -90.00 | -89.19 | 0.81 | -0.35 | 0.52 | 33.14 | 4.45 | 1.9 |
| -108.00 | -109.65 | -26.56 | -26.08 | 1.56 | 13.07 | 36.60 | 14.17 | 4.33 | 0.9 |
| 72.00 | 70.02 | 26.56 | 26.42 | 1.78 | -13.21 | -36.33 | -14.31 | 4.32 | 1.0 |
| 180.00 | 177.92 | -26.56 | 23.73 | 3.40 | 38.54 | -1.40 | 12.63 | 4.21 | 3.5 |
| 0.00 | -2.21 | 26.56 | 25.19 | 2.42 | -38.16 | 1.47 | -13.38 | 4.22 | 3.4 |

## V. Conclusions

A method for estimating the magnetic dipole moment of a coil fed by a sinusoidal current was presented, and validated both by means of simulations and experimentally. The proposed approach is efficient, requiring only 3 measurements, and achieves a good accuracy, with a relative error of less than 4% on the magnitude of the magnetic dipole moment and less than 4° on its bearing.

## Acknowledgment

This research activity was funded through grant PRIN 2015C37B25 by the Italian Ministry of Instruction, University and Research (MIUR), whose support the authors gratefully acknowledge.

## References.